\documentclass[]{krakproc}
\usepackage{graphicx}

\def\bmath#1{{\bf #1}}

\def\eqref#1{Eq.~(\ref{#1})}

\def\eqsdash#1#2{Eqs.~(\ref{#1}-\ref{#2})}
\def\Eqref#1{Eq.~(\ref{#1})}

\def\Eqsdash#1#2{Eqs.~(\ref{#1}-\ref{#2})}
\def\Eqsand#1#2{Eqs.~(\ref{#1}) and~(\ref{#2})}
\def\figref#1{Fig.~\ref{#1}}

\def\const{{\rm const}}

\def\bea{\begin{eqnarray}}
\def\eea{\end{eqnarray}}

\def\({\left(}
\def\){\right)}
\def\<{\langle}
\def\>{\rangle}
\def\lt{\left}
\def\rt{\right}
\def\bl{\bigl}
\def\br{\bigr}

\def\dd{\partial}
\def\diff{d}

\def\vdel{\bmath{\nabla}}
\def\delpar{\nabla_{\parallel}}

\def\vperp{v_\perp}

\def\vu{\bmath{u}}

\def\vB{\bmath{B}}

\def\vb{\bmath{\hat b}}

\def\pperp{p_\perp}
\def\ppar{p_\parallel}
\def\Tperp{T_\perp}
\def\Tpar{T_\parallel}

\def\Brms{B_{\rm rms}}
\def\Bbend{B_{\rm bend}}
\def\Kbend{K_{\rm bend}}

\def\rbend{\rho_{i,{\rm bend}}}

\def\ud{u_\nu}
\def\vth{v_{\rm th}}

\def\lpar{l_{\parallel}}

\def\ld{l_{\nu}}
\def\lf{l_0}
\def\lres{l_{\eta}}
\def\mfp{\lambda_{\rm mfp}}
\def\lB{l_B}

\def\Bsq{{\langle B^2 \rangle}}

\def\nui{\nu_{ii}}

\def\nuB{\nu}

\def\vk{\bmath{k}}
\def\kpar{k_{\parallel}}
\def\kperp{k_{\perp}}
\def\vkperp{\vk_{\perp}}

\def\dvu{\delta\vu} 
\def\dvuperp{\delta\vu_\perp} 
\def\dupar{\delta u_\parallel}

\def\dvB{\delta\vB}

\def\dB{\delta B}
\def\dvb{\delta\vb} 
\def\dpperp{\delta\pperp}
\def\dppar{\delta\ppar}

\def\vA{v_A}
\def\gmax{\gamma_{\rm max}}

\def\Re{{\rm Re}}

\def\Pm{{\rm Pm}}

\begin{document}

\title{Magnetised plasma turbulence in clusters of galaxies}
\author{A.\ A.\ Schekochihin,$^1$ S.\ C.\ Cowley,$^{2,3}$ 
R.\ M.\ Kulsrud,$^4$\\ G.\ W.\ Hammett$^4$ and P.\ Sharma$^4$}
\institute{$^1$DAMTP, University of Cambridge, Cambridge CB3 0WA, UK;
as629@damtp.cam.ac.uk\\
$^2$Department of Physics and Astronomy, UCLA, Los Angeles, 
California 90095-1547, USA\\
$^3$Plasma Physics, Imperial College London, London~SW7~2BW, UK\\
$^4$Plasma Physics Laboratory, Princeton University, Princeton, NJ 08543-0451, USA}
\markboth{A.A. Schekochihin et al.}{Magnetised plasma turbulence in clusters of galaxies}

\maketitle

\begin{abstract}
Cluster plasmas are magnetised already at very low 
magnetic field strength. 
Low collisionality implies that conservation 
of the first adiabatic invariant results in an anisotropic viscous 
stress (Braginskii viscosity) or, equivalently, anisotropic plasma pressure. 
This triggers firehose and mirror instabilities, 
which have growth rates proportional to the wavenumber 
down to scales of the order of ion Larmor radius. 
This means that MHD equations with Braginskii viscosity 
are not well posed and fully kinetic description is necessary. 
In this paper, we review the basic picture of small-scale 
dynamo in the cluster plasma and attempt to reconcile it 
with the existence of plasma instabilities at collisionless scales. 
\end{abstract}

The intracluster medium (ICM) is a unique environment 
for studying astrophysical turbulence: it is made of amorphous 
diffuse fully ionised plasma with almost no net 
rotation or shear. The ICM is, therefore, in a certain 
sense, a case of ``pure'' turbulence. The presence of magnetic 
fields in the ICM with $B\sim1...10~\mu$G 
and field scale $\sim 1$~kpc is well documented 
(e.g., \cite{Carilli_Taylor}). 
Because of the availability of the rotation measure data 
from extended radio sources, it is possible 
to infer the magnetic-energy spectra for some of the 
clusters with spatial resolution of $\sim 0.1$~kpc 
(\cite{Vogt_Ensslin,Vogt_Dolag_Ensslin}). 
This is due to improve dramatically with the arrival 
of LOFAR and, especially, SKA (\cite{Gaensler_Beck_Feretti}). 
At the same time, there is a rising interest among astronomers 
in measuring the velocity spectra of the cluster turbulence 
using instruments both future (ASTRO-E2, \cite{Sunyaev_Norman_Bryan}) 
and present (XMM-Newton, \cite{Schuecker_etal}). 
Once we have both kinetic and magnetic-energy spectra, 
a clear observational picture of the magnetised cluster 
turbulence will emerge. The purpose of this paper is to 
review our theoretical understanding of this turbulence. 

There is no consensus on the exact mechanism that 
drives turbulence in clusters. The main candidates are 
galaxy wakes, jets, and cluster mergers. We will leave 
a discussion of these outside the scope of this paper. 
The important point for us is that all these mechanisms 
imply the outer scale of the cluster turbulence 
in the range $\lf\sim10^2...10^3$~kpc with outer scale 
velocities comparable to the thermal (sound) speed 
$\vth=(T/m_i)^{1/2}\sim10^3$~km/s, where $m_i$ is ion mass 
(assuming hydrogen plasma) and $T\sim10^8$~K the plasma temperature 
(assuming for simplicity $T_i=T_e$). 
This gives the outer eddy timescale of order $\lf/\vth\sim10^9$~yr. 

The ion kinematic viscosity is $\nu\sim\vth^2/\nui$, 
where $\nui=4\pi n e^4\ln\Lambda m_i^{-1/2}T^{-3/2}$ is the ion-ion 
collision frequency, $n\sim10^{-2}...10^{-3}$~cm$^{-3}$ is the ion 
number density (for cluster cores), $e$~is the electron charge, 
and $\ln\Lambda\sim20$ is the Coulomb logarithm 
(\cite{Spitzer_book}). 
This gives Reynolds numbers in the range $\Re\sim10^2...10^3$. 
As the turbulent velocities at the outer scales are 
$\sim\vth$, the turbulence in the inertial range will be 
subsonic. It is natural to assume that Kolmogorov's dimensional 
theory should apply at least approximately. Then the 
viscous scale of the turbulence is $\ld\sim\lf\Re^{-3/4}\sim10...30$~kpc. 
These numbers appear to agree quite well with observations 
of turbulence in the Coma cluster via pressure maps (\cite{Schuecker_etal}). 

Because the ICM plasma is fully ionised, the viscous cutoff is, 
in fact, quite different from the usual Laplacian viscosity 
that normally appears in MHD equations. 
It can be shown that at frequencies below the ion cyclotron 
frequency~$\Omega_i=eB/cm_i$ 
and scales above the ion Larmor radius~$\rho_i=\vth/\Omega_i$, 
fluid velocity and magnetic field satisfy 
\bea
\label{eq_u}
\rho\,{\diff\vu\over\diff t}
&=& -\vdel\(\pperp + {B^2\over8\pi}\) 
+ \vdel\cdot\lt[\vb\vb(\pperp-\ppar)\rt]
+ {\vB\cdot\vdel\vB\over4\pi},\\
\label{eq_B}
{\diff\vB\over\diff t} 
&=& \vB\cdot\vdel\vu - \vB\vdel\cdot\vu 
+ \eta\nabla^2\vB, 
\eea
where $\diff/\diff t = \dd/\dd t + \vu\cdot\vdel$, 
$\rho=m_i n$ is mass density, 
$\vb=\vB/B$, 
$\pperp$ and $\ppar$ are the plasma 
pressures perpendicular and parallel to the magnetic field, 
$\eta\sim T^{-3/2}m_e^{1/2} e^2 c^2\ln\Lambda/4\pi$ is 
the magnetic diffusivity (\cite{Spitzer_book}; we ignore 
the order-unity difference between $\eta_\parallel$ and $\eta_\perp$). 
\Eqref{eq_u} 
is the appropriate description for the fluid motions in the ICM 
provided the ions are magnetised, i.e., $\rho_i\ll\mfp$, 
where $\mfp\sim\vth/\nui\sim1...10$~kpc is the ion 
mean free path. This requirement 
is satisfied if $B\gg10^{-18}$~G, which is far below 
the observed field strengths of $1...10~\mu$G. 
Note that $10~\mu$G is roughly the field strength 
corresponding to the energy of the viscous-scale eddies 
($\sim\rho\vth^2\Re^{-1/2}$), which is the lower bound 
for dynamically important fields. 
Thus, even for dynamically 
weak fields, the plasma is already very well magnetised. 

The plasma pressures in \eqref{eq_u} must be calculated 
kinetically. At low frequencies ($\omega\ll k\vth$) 
and at scales above $\mfp$, 
the total isotropic pressure $\pperp+B^2/8\pi$ can be 
determined from incompressibility, $\vdel\cdot\vu=0$, 
while the pressure anisotropy is calculated perturbatively 
from the kinetic theory (\cite{Braginskii}):
\bea
\pperp-\ppar = 3\rho\nuB\,\vb\vb:\vdel\vu \equiv \rho\vth^2\Delta, 
\label{dp_brag}
\eea
where $\nuB\sim\vth\mfp\sim\vth^2/\nui$ is the ion viscosity 
and $\Delta$ is the dimensionless measure of the 
pressure anisotropy. The pressure anisotropy 
is proportional to (the parallel component of) the 
rate of strain, so it is determined by the viscous-scale 
eddies' turnover rate $\ud/\ld$. 
For Kolmogorov turbulence 
with outer-scale velocity~$\sim\vth$, we have
$\ud/\vth\sim\Re^{-1/4}$ and $\ld/\mfp\sim\Re^{-1/4}$, 
so $\Delta\sim\ud\mfp/\vth\ld\sim\Re^{-1/2}$. 

\Eqsdash{eq_u}{dp_brag} plus the incompressibility  
condition $\vdel\cdot\vu=0$ are a closed system that 
we will call the Braginskii MHD. Two key properties of these equations 
should be noted. First, velocity gradients transverse to the magnetic field 
(e.g., shear-Alfv\'en-polarised fluctuations) are not dissipated. 
Velocity fluctuations can now penetrate below the viscous cutoff. 
Second, the velocities that are dissipated 
are those that change the strength of the magnetic field: 
indeed, from \eqref{eq_B}, 
\bea
\label{eq_lnB}
{1\over B}{\diff B\over\diff t} = \vb\vb:\vdel\vu. 
\eea
Physically, the Braginskii viscosity 
encodes a fundamental property of 
charged particles moving in a magnetic field: the conservation 
of the first adiabatic invariant $\mu=m\vperp^2/2B$, 
which is only weakly broken when $\mfp\gg\rho_i$.
This means that any change in $B$ must be accompanied by 
a proportional change in $\pperp$. 
Therefore, the emergence of the pressure anisotropy is a natural 
consequence of the changes in the magnetic-field strength and 
vice versa. 

Let us ignore for a moment the inability of the Braginskii 
viscosity to damp all velocity fluctuations and consider 
the turbulent motions between the outer scale and the 
viscous scale and the random stretching of the magnetic field 
by these motions, or the small-scale dynamo.  
Such a view might appear justified because 
the Braginskii viscosity cuts off any motions that 
change the field strength, so  the small-scale dynamo action 
is associated only with velocities above the viscous scale. 
The turbulent small-scale dynamo is a problem 
that we have studied extensively 
(for the full account of our work and a list 
of relevant references, see \cite{SCTMM_stokes}). 
Here we reiterate briefly the main points. 

In the weak-field regime (no back reaction on the flow), 
the turbulent stretching is done primarily by the 
viscous-scale eddies because they turn over the fastest. 
The magnetic energy grows exponentially roughly at their 
turnover rate $\ud/\ld\sim10^{-8}~{\rm yr}^{-1}$. 
%Regardless of the initial field configuration, 
The turbulence tends to arrange the magnetic fields 
in long thin flux sheets (or ribbons): the field reverses 
its direction at the smallest scale $\lB$ available to it 
(if the magnetic cutoff is modelled by the Ohmic diffusivity $\eta$, 
it is the resistive scale: 
$\lB\sim\lres\sim\ld\Pm^{-1/2}$, where $\Pm=\nu/\eta$) 
but field lines curve at the scale 
of the flow ($\lpar\sim\ld$, the viscous scale) 
except in the sharp bends of the folds. The field is in the state 
of reduced tension everywhere: 
both in straight and bent parts of the folds,  
$BK^{1/2}\sim\const$, 
where $K=|\vb\cdot\vdel\vb|$ is the field-line curvature. 
Thus, the field strength and the field-line curvature are 
anticorrelated. 

The tendency to fold the field is a kinematic property of random stretching. 
A key numerical finding is that the folded structure 
is also a feature of the nonlinear saturated state of the dynamo. 
The folded structure allows the small-scale 
direction-reversing magnetic field to back react 
on the flow in a spatially coherent way. The back-reaction 
mechanism we have proposed consists in
the velocity gradients becoming locally anisotropic with respect 
to the direction of the folds (with $\vb\vb:\vdel\vu$ partially 
suppressed) so that the dynamo saturates at the marginally stable 
balance between reduced ``parallel'' stretching and 
``perpendicular'' mixing. 

In the kinematic regime, the characteristic fold length 
is the viscous scale, $\lpar\sim\ld$. 
In the saturated state, the folds 
tend to elongate to the outer scale, $\lpar\sim\lf$. 
We have proposed that the saturation is controlled 
by the anisotropisation of the outer-scale eddies 
by the folded fields, while the inertial range ($\lf>l>\ld$) 
is populated by Alfv\'en waves that propagate along the folds 
with the dispersion relation $\omega^2=\vk\vk:\vb\vb\Bsq/4\pi\rho$. 

The physical considerations outlined above provide 
a qualitative framework for interpreting 
the saturated spectra of magnetic and kinetic energies 
that are seen in the numerical simulations 
of forced isotropic MHD turbulence: magnetic energy 
significantly exceeds kinetic energy at small scales and, in the 
case of large $\Pm$, tends to concentrate below the viscous 
scale (see, e.g., \cite{SCTMM_stokes,HBD_pre}). 
Such a magnetic spectrum makes sense if it is thought of as a superposition 
of folds with field reversals at the resistive scale and 
and Alfv\'en waves propagating along the folds. 

Recently published magnetic-energy spectra based 
on RM measurements from extended radio sources 
for Abell~400, 2255, 2634, and Hydra~A clusters 
(see \cite{Vogt_Ensslin,Vogt_Dolag_Ensslin}, and En{\ss}lin's contribution 
in these Proceedings) show magnetic energy to be peaked at $\sim1$~kpc 
with the overall shape of the spectra quite 
similar to those that emerge in numerical simulations cited above. 
Does this mean that these simulations correctly describe 
cluster turbulence? The answer to this question 
is probably in the negative. All existing 
simulations either use isotropic viscosity and 
Ohmic diffusivity to model kinetic and magnetic cutoffs 
or rely on numerical dissipation.
Clearly, simulations with a magnetic cutoff at or not far below 1~kpc 
(i.e., about a decade below the viscous scale) are likely 
to produce magnetic spectra that will look roughly 
like the observed ones. 
However, if we estimate the resistive cutoff scale based on 
the formula for Ohmic diffusivity given above, we will get 
$\lres\sim10^4$~km --- many orders of magnitude below 1~kpc. 
Furthermore, no published numerical study has 
included the effect of anisotropic viscosity to address 
the undamped velocity fluctuations at subviscous scales. 
In what follows, we show that the Braginskii 
viscosity not only fails to efficiently damp kinetic energy 
at the viscous scale but triggers 
fast-growing instabilities at subviscous scales. 
This fundamentally alters the structure of the turbulence at these scales. 

It has, in fact, been known for a long time that pressure anisotropy 
leads to plasma instabilities 
(\cite{Rosenbluth,Chandrasekhar_Kaufman_Watson,Parker1,Vedenov_Sagdeev}). 
A vast literature exists on these instabilities, which we 
do not attempt to review. The standard approach 
is to postulate a bi-Maxwellian equilibrium 
distribution with $\Tperp\neq\Tpar$ 
(see, e.g., \cite{Ferriere_Andre} and references therein). 
We do not need to adopt such a description because 
\eqsdash{eq_u}{dp_brag} 
incorporate the pressure anisotropy in a self-consistent way. 

Consider a turbulent cascade that originates from 
the large-scale driving, extends down to the viscous scale, and 
gives rise to velocity and magnetic fields $\vu$ and $\vB$ that vary 
on time scales $\sim(\nabla u)^{-1}$ and on spatial scales $\sim\ld$. 
We study the stability of such fields. 
The presence of turbulent shear (velocity gradients)
gives rise to the pressure anisotropy given by \eqref{dp_brag}. 
We now look for linear perturbations $\dvu$, $\dvB$, 
$\dpperp$, $\dppar$ that have frequencies $\omega\gg\nabla u$,
and wavenumbers $k\gg\ld^{-1}$. 
With respect to these perturbations, the unperturbed 
rate-of-strain tensor $\vdel\vu$ can be viewed as constant in space 
and time. Linearising \eqref{eq_u} and neglecting temporal and 
spatial derivatives of the unperturbed quantities, we get
\bea
\nonumber
-i\omega\rho\dvu &=& -i\vk\,\delta\(\pperp + {B^2\over8\pi}\) + 
\(\pperp-\ppar+{B^2\over4\pi}\)\delta\bl(\vb\cdot\vdel\vb\br)\\ 
&&+\,\,\vb\,\delta\lt[\delpar\(\pperp-\ppar\)
-\(\pperp-\ppar - {B^2\over4\pi}\){\delpar B\over B}\rt]\qquad
\label{eq_u_lin} 
\eea
and, from \eqref{eq_B}, $\dvb=-(\kpar/\omega)\dvuperp$ 
and $\delta\bl(\vb\cdot\vdel\vb\br)=i\kpar\dvb$. We have denoted  
$\delpar=\vb\cdot\vdel$ and $\kpar=\vb\cdot\vk$. 
In the resulting dispersion relation, it is always 
possible to split off the part that corresponds to 
the modes that have shear-Alfv\'en-wave polarisation, 
$\dupar=0$, $\vkperp\cdot\dvuperp=0$: 
\bea
\label{dr_shear}
\omega^2 = \kpar^2\vth^2\(\Delta + 2\beta^{-1}\), 
\eea
where $\beta=2\vth^2/\vA^2=8\pi\rho\vth^2/B^2$.
When the magnetic energy is larger than the energy of the 
viscous-scale eddies, $\beta\ll|\Delta|^{-1}\sim\Re^{1/2}$, 
\eqref{dr_shear} describes shear Alfv\'en waves. 
In the opposite limit, $\beta\gg|\Delta|^{-1}$, an instability appears: 
when $\Delta<0$, the growth rate is 
$\gamma=\kpar\vth|\Delta|^{1/2}\sim\kpar\vth\Re^{-1/4}$. 
Since $\kpar\gg\ld^{-1}$, the instability is faster than the rate of 
strain, $\gamma\gg\nabla u\sim\vth\Re^{-1/4}/\ld$, in 
accordance with the assumption made in the derivation of \eqref{dr_shear}. 
This instability, triggered by $\ppar>\pperp$, 
is called the firehose instability. 
It does not depend on the way the pressure perturbations are calculated 
because it arises from the perturbation of the field-line curvature 
$\vb\cdot\vdel\vb$ in \eqref{eq_u_lin}: to linear order, 
it entails no perturbation of the field strength 
[from \eqref{eq_B}, $\dB/B=-(\kpar/\omega)\dupar$] 
and, therefore, does not alter the pressure. 

In order to determine the stability 
of perturbations with other polarisations,
$\dpperp$ and $\dppar$ have to be computed. 
We consider scales smaller than 
the mean free path, $k\mfp\gg1$, where the plasma is 
collisionless. 
It is sufficient to look for ``subsonic'' perturbations such 
that $\omega\ll k\vth$ because the high-frequency perturbations 
are subject to strong collisionless damping and cannot 
be rendered unstable by a small anisotropy. 
A straightforward linear kinetic calculation 
(\cite{SCKHS_brag}) shows that the slow-wave-polarised 
perturbations are unstable with the growth rate 
\bea
\gamma = \(2\over\pi\)^{1/2}|\kpar|\vth\lt[ 
\Delta\(1-{\kpar^2\over2\kperp^2}\) 
- \beta^{-1}\(1+{\kpar^2\over\kperp^2}\)\rt].
\label{gamma_mirror}
\eea 
Modes with $\kpar>\sqrt{2}\kperp$ are unstable if 
$\Delta<0$ (slow-wave-polarised firehose instability); 
modes with $\kperp>\kpar/\sqrt{2}$ are unstable if 
$\Delta>0$ (mirror instability). 

The instabilities we have described have growth rates 
$\propto\kpar$. Thus, {\em MHD equations with Braginskii 
viscosity do not constitute a well-posed problem.} This means, 
for example, that any numerical simulation of these equations 
will blow up at the grid scale unless additional 
isotropic viscosity is introduced (this has 
been confirmed by J.~L.~Maron, unpublished). 

\Eqsand{dr_shear}{gamma_mirror} are valid in the limit $k\rho_i\to0$. 
If finite $\rho_i$ is taken into account, 
the growth rates peak at $k\sim\rho_i^{-1}$ with 
$\gmax\sim|\Delta|\Omega_i$ for the mirror 
and $\gmax\sim|\Delta|^{1/2}\Omega_i$ 
for the shear-Alfv\'en-polarised firehose. 
For clusters, this gives values of $\gmax$ in the range 
$10^{-8}...10^{-7}(B/10^{-18}~{\rm G})~{\rm yr}^{-1}$ 
--- much faster than the viscous-eddy turnover rate.  

We have shown that, given sufficiently high $\beta$, 
the firehose and mirror instabilities occur in the regions 
of decreasing ($\Delta<0$) and increasing ($\Delta>0$) 
magnetic-field strength, respectively. 
The small-scale dynamo action by the 
viscous-scale motions associated with the turbulent 
cascade from large scales will produce regions of both types: 
the folded structure explained above contains 
growing straight direction-reversing fields 
and weakening curved fields in the corners of the folds. 
This structure is intrinsically unstable: 
straight growing fields to the mirror, 
curved weakening fields to the firehose instability.
It seems plausible, however, that the magnetic field does 
grow until $\beta\sim|\Delta|^{-1}\sim\Re^{-1/2}$, 
so the instabilities are stabilised. 
This corresponds to $B\sim10~\mu$G, which is
comparable to the observed field strength in clusters. 

After the instabilities are quenched, 
the fluid motions can be separated into two classes: 
above the viscous scale, there is the usual turbulent 
cascade, to which the picture of small-scale 
dynamo given in \cite{SCTMM_stokes} and 
outlined above applies; below the viscous scale, 
the slow-wave-polarised motions are damped, while the 
shear Alfv\'en waves (the stable counterpart of the 
firehose instability) can exist. If we conjecture
the survival of the folded structure, 
which is a generic outcome of random stretching, then 
the Alfv\'en waves will propagate along the folds. 
Note that the linear stability analysis outlined 
above can be recast in terms of the perturbations of 
striped direction-reversing fields represented by the tensor $\vB\vB$. 
%as long as the reversal scale 
%is smaller than the scale of the perturbations we are considering. 
What we still lack is an estimate of the reversal scale. 

\begin{figure}[t!]
\centerline{
\resizebox{3in}{!}{
\includegraphics{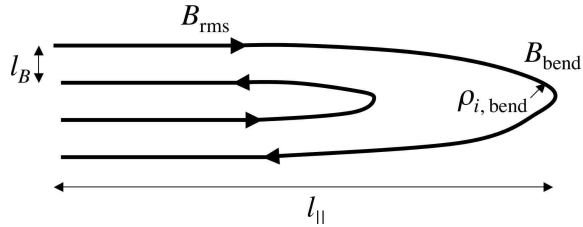}}
}
\caption{A sketch of the folded structure and 
scales referred to in the derivation of \eqref{lB_estimate}.}
\label{fig_fold}
\end{figure}

Even for rms field strengths sufficient 
to quench the instabilities, there will always be regions 
where the field is weak. These regions can become unstable unless 
they are smaller than the local ion Larmor radius. 
We conjecture that the overall structure of the magnetic field 
is determined by the requirement that the field scale in 
the weak-field regions should be smaller than local 
value of $\rho_i$. 
The following rather tentative argument is an attempt
to estimate the characteristic field scale based on this idea. 
As we mentioned above, the folded structure is characterised 
by the anticorrelation between the field strength 
and field-line curvature: $BK^{1/2}\sim\const$. 
The weak-field regions are the bending 
regions where the field is curved (see \figref{fig_fold}). 
A simple flux-conservation argument gives 
$\Bbend/\Brms\sim\lB/\lpar$, where $\Brms$ is 
the field in the strong-field regions, 
$\lB$ is the reversal scale, 
and $\lpar$ is the fold length, which, in the nonlinear regime, 
can be as long as the outer scale $\lf$ of the turbulence. 
We estimate the curvature in the bending regions 
by $\Kbend\sim1/\rbend$, 
where $\rbend=\rho_i\Brms/\Bbend$ is the ion Larmor radius 
in the bending regions 
and $\rho_i=\vth/\Omega_i=cm_i^{1/2}T^{1/2}/e\Brms$ 
is the ion Larmor radius in the strong-field regions. 
The curvature in the strong-field regions is $\sim1/\lpar$. 
Then $\Bbend/\Brms\sim(\rbend/\lpar)^{1/2}$. 
Assembling these relations and substituting numbers, 
we obtain the following estimate 
\bea
\lB\sim \rho_i^{1/3}\lpar^{2/3}
\sim 10^{-2} \({T\over 10^8~{\rm K}}\)^{1/6}
\({\Brms\over 1~\mu{\rm G}}\)^{-1/3} 
\({\lpar\over 1~{\rm Mpc}}\)^{2/3}{\rm kpc}.
\label{lB_estimate}
\eea
This is still substantially smaller than the scale at which 
the observed spectrum peaks, but not 
entirely unreasonable, considering the simplistic nature of our argument. 
Note that the shear Alfv\'en waves can exist at scales below $\lB$, 
with a cutoff at $\kperp\rho_i\sim1$, so 
the turbulence spectra are likely to have extended power tails. 

A somewhat more detailed exposition can be found in \cite{SCKHS_brag}.\\

A.A.S.\ was supported by the UKAFF Fellowship. 
This work was supported in part by the NSF Grant No.~AST~00-98670 
and by the US DOE Center for Mutiscale Plasma Dynamics. 
G.W.H.\ and P.S.\ were supported by the US DOE 
Contract No.~DE-AC02-76CH03073. 

\end{document}